# Representation in terms of displaced number states and realization of elementary linear operators based on it

Sergey A. Podoshvedov

*Department of General and Theoretical Physics, South Ural State University, Lenin Av. 76, Chelyabinsk, Russia*

We develop a new method of representation of quantum states in terms of the displaced number (Fock) states. We call the decomposition $\alpha-$ representation, where $\alpha$ is an amplitude of the base displaced states. Analytical expressions of the $\alpha-$ representation for the displaced number states, two-mode squeezed vacuum (TMSV) and superposition of vacuum and single photon are obtained. New mechanism of manipulations with quantum information based on symmetry properties of the wave amplitudes relatively sign of the displacement amplitude in $\alpha-$ representation of the state is proposed. Extraction of the displaced single photons from the target state is employed for building of elementary linear operators: two-qubit control-sign gate and Hadamard matrix. Entangled coherent state shifts two-mode target qubit on phase plane by different values. Registration of the single photon in auxiliary mode results in constructive interference of the control and target qubits. The results are exact, any approximations are not exploited. We show implementation of the gates with almost unity fidelity of the output states in realistic scenario.
.

e-mail: sapo66@mail.ru

## 1. Introduction

Quantum information with qubits as carriers provides a totally new way of information processing to execute intriguing tasks which would not be possible by means of traditional classical methods [1]. Quantum parallelism and entanglement are fundamental features of quantum mechanics which provide a significant speed-up for certain computational tasks such as factorizing a large integer [2] and unsorted database searching [3]. Qubits can physically be realized in optical systems. All-optical quantum computation has its own advantages. Photons can be easily manipulated with linear optical devices at room temperature, they propagate very fast, they can resist to influence of environment and they preserve coherence in free space long when they do not interact with medium. It was demonstrated in [4] (KLM proposal) that, given single-photon sources and high-efficiency detectors, quantum computation can be implemented by linear optics alone. Practically, this method is complex and unlikely effective since tens thousands of operations per near deterministic entangling gate are required. Practical problem of the approach is inefficiency of the single photon sources and photo-detection.

Measurement-based linear optical quantum operators look attractive with theoretical point of view [5]. Universal quantum operations are produced by measurement of the particles in corresponding basis provided that large entangled cluster state is prepared beforehand [6]. But practical production of the cluster states is hardly possible. So proposal in [7] needs medium only with nonlinearity responsible for cross-phase modulation that maybe not feasible. So, optical fiber has cubic nonlinearity that contributes to comparable self- and cross modulation of the propagating optical fields.

In KLM paradigm single photons are treated as qubits. But logical qubit can be represented as a superposition of two coherent states with the same amplitude but opposite phases. As rule, one says about even and odd superposition of coherent states (SCSs) in dependency on phase relation between components of the superposition. The SCSs are orthogonal, can be recognized as optical analogue of the cat states [8] and can be exploited for quantum information processing [9]. Single qubit gates are required to control qubit locally. Strong non-linear interaction [10] to realize Hadamard gate with coherent states, key element of the single qubit transformations, must be employed. Nonlinear effects in existing materials are extremely small and not enough [11]. In Ref. [12] a qubit rotation and a nontrivial control-NOT gate are studied by applying phase-space displacements and beam splitters followed by detectors. The gates are probabilistic and successfully work when definite outcome of projective or POVM (Positive Operator-Valued Measure) [1] measurements in auxiliary modes is observed. Experimental Hadamard gate was demonstrated in [13]. Beam splitter with high transmittance provides an alternative mechanism (subtraction of photons) to achieve effective nonlinear effect on input state when reflected mode is measured. Such measurement-based technique is realized in many practical proposals [14-21].

Big things are classical and small stuff like atoms and photons is quantum. One of the aims of quantum investigations is to push further into the macroscopic world to have controllable quantum effects in larger systems. The large quantum systems may behave in ways that are hard to anticipate. To succeed in pushing quantum effects up to the macroscopic scale, we need to overcome decoherence effect. Decoherence is responsible for difference between the macroscopic and microscopic objects. Decoherence may be serious obstacle for KLM [4] and cluster state [6] approaches. The quantum system with larger number of particles may be subjected more influence of decoherence since every particle may inevitably interact with the environment amplifying the total effect. It causes the system with large number of quantum particles may start to behave classically after some short time after its quantum birth. If to expend a lot of efforts over time, we may finally succeed in beating decoherence effect in multi-particle quantum systems. But it might happen the goal cannot be reached ever because it is impossible.

We can to look at a problem of physical system to be both macroscopic and quantum under other angle. We make use of displaced version of number states of quantum oscillator [22]. A coherent state has the mean non-zero complex amplitude with equal uncertainties in the two quadrature phases as a vacuum. The coherent state or the same displaced vacuum can be represented by an "error circle" in a complex plane whose axes are quadrature components [23]. The displaced versions of the number states with large displacement amplitude also leave some images on phase plane with center far away from origin of coordinates. They may be considered as macroscopic in the case and conserve their quantum properties. The displaced number states may contain infinite number of elementary particles distributed over some law. Nevertheless, we can also work with the displaced number state as with one "large" particle neglecting its internal structure. This implies equality of sets of number and displaced number states, for example, in application to quantum information processing problems. We only need to connect the sets of the states by unitary matrix corresponding to the displacement operator.

Here, we derive transformation matrix between arbitrary sets of the displaced number states with different amplitudes of displacement. It enables to introduce new representation of arbitrary state. Analytical expressions of the representation of some states are presented. Given the representation of the states, new method of generation of the target state with help of extraction of displaced number state from correlated part of composite system [21] is developed. It was experimentally shown in [24] that use of squeezed vacuum and photon subtraction technique combined with displacement operator generates arbitrary superposition of squeezed vacuum and squeezed single photon with high precision. We develop general treatment of construction of elementary single- and two-qubit controlled gates based on extraction of the displaced state. Earlier the displaced single photon is proposed to use for dense coding protocol [25] and quantum key distribution with displaced states [26] as carries of information.

## 2. $\alpha -$ representation of pure states

Quantization of the electromagnetic field is accomplished by choosing classical complex amplitudes to be mutually adjoint operators. The dynamical behavior of the electric-field amplitudes is described by an ensemble of independent harmonic oscillators obeying well-known communication relations. The pure state in each mode can be described by some superposition of the number (Fock) states of the harmonic oscillator in Hilbert space appropriate to that mode. The number states are the partial case of the displaced number ones with amplitude of displacement $\alpha$ equated to zero. Moreover, the number and displaced number states are connected with each other by unitary transformation. Therefore, any pure state can be also represented in terms of the displaced number states. We are going to call the decomposition over displaced number states $\alpha -$ representation of the state. $\alpha -$ representation can be also introduced for the mixed states but it is out of the consideration. Three states (displaced number state, TMSV and superposition of vacuum and single photon) are chosen as exemplary to derive their $\alpha -$ representations.

### 2.1 Transformation matrix for displacement operator

The eigenvalues of Hamiltonian of the quantum harmonic oscillator are defined by integers $(n = 0,1,2,...,\infty)$. Corresponding eigenstates $|n\rangle$ are known as number or Fock states. The number (Fock) states are orthogonal $\langle n|m\rangle = \delta_{nm}$, where $\delta_{nm} = 1$, if $n = m$, otherwise $\delta_{nm} = 0$. Displaced number states are obtained by additional application of the displacement operator [23]

$$D(\alpha) = \exp(\alpha a^+ - \alpha^* a),\tag{1}$$

to the number state $|n\rangle$ [20-22]

$$|n,\alpha\rangle = D(\alpha)|n\rangle,\tag{2}$$

where $\alpha$ is an amplitude of the displacement and $a$, $a^+$ are the bosonic annihilation and creation operators. Set of the displaced number states is complete for arbitrary value of $\alpha$. Annihilation and creation operators $A = a - \alpha$ and $A^+ = a^+ - \alpha^*$, respectively, can be introduced. Their application to the displaced number stare $|n,\alpha\rangle$ yields

$$A|n,\alpha\rangle = \sqrt{n}|n-1,\alpha\rangle,\tag{3}$$

$$A^+|n,\alpha\rangle = \sqrt{n+1}|n+1,\alpha\rangle.\tag{4}$$

The mean energy of the displaced number state in non-dimensional units (neglecting vacuum fluctuations)

$$\langle n,\alpha|A^+A|n,\alpha\rangle = n + |\alpha|^2,\tag{5}$$

becomes the sum of the number state (quantum) energy $n$ and classical wave intensity $|\alpha|^2$. This means the displaced number states are not physically similar to the number states. Although, they have some quantum noise properties in common [23]. The displaced Fock states (1) are defined by two numbers: quantum discrete number $n$ and classical continuous parameter $\alpha$. Somewhat, it may testify about manifestation of particle-wave duality of the states, especially, of the coherent state or the same displaced vacuum.

Choose two sets of orthogonal displaced number states

$$\left\{n,\alpha\right), n = 0,1,2,...,\infty\right\}, \tag{6}$$

$$\left\{n,\alpha'\right), n = 0,1,2,...,\infty\right\}, \tag{7}$$

where, in general case, $\alpha \neq \alpha'$. Every element from one set can be expressed through states from another for its completeness. Coefficients of the decomposition are the inner product [27].

$$\left\langle n,\alpha' \middle| n,\alpha \right\rangle \tag{8}$$

which takes as input two vectors $|n,\alpha\rangle$ and $|n,\alpha'\rangle$ from sets (6, 7) and produces, in general case, a complex number as output [1].

Let us present our derivation of the $\alpha-$ representation of the displaced number states. There exists the following chain of transformations with operators and states

$$|l,\alpha'\rangle = D(\alpha')|l\rangle = \exp\left(-\left|\alpha'\right|^2 \middle/ 2\right)\exp\left(\alpha' a^+\right)\exp\left(-\alpha'^* a\right)|l\rangle =$$

$$\exp\left(-\left|\alpha'\right|^2 \middle/ 2\right)\exp\left(\beta a^+\right)\exp\left(\alpha a^+\right)\exp\left(-\alpha^* a\right)\exp\left(\alpha^* a\right)\exp\left(-\alpha' a\right)|l\rangle =$$

$$\exp\left(-\left|\alpha'\right|^2 \middle/ 2\right)\exp\left(\beta a^+\right)\exp\left(\alpha a^+\right)\exp\left(-\alpha^* a\right)\exp\left(\left(\alpha-\alpha'\right)^* a\right)|l\rangle =$$

$$\exp\left(-\left(\left|\alpha'\right|^2 - \left|\alpha\right|^2\right)\middle/ 2\right)\exp\left(\beta a^+\right)\exp\left(-\left|\alpha\right|^2 \middle/ 2\right)\exp\left(\alpha a^+\right)\exp\left(-\alpha^* a\right)\exp\left(\left(\alpha-\alpha'\right)^* a\right)|l\rangle$$

$$= \exp\left(-\left(\left|\alpha'\right|^2 - \left|\alpha\right|^2\right)\middle/ 2\right)\exp\left(\beta a^+\right)D(\alpha)\exp\left(\left(\alpha-\alpha'\right)^* a\right)|l\rangle =$$

$$\exp\left(-\left(\left|\alpha'\right|^2 - \left|\alpha\right|^2\right)\middle/ 2\right)\exp\left(\beta a^+\right)D(\alpha)$$

$$\left(1 + \left(\alpha-\alpha'\right)^* a + \left(\alpha-\alpha'\right)^{*2} a^2 \middle/ 2! + \left(\alpha-\alpha'\right)^{*3} a^3 \middle/ 3! + ... + \left(\alpha-\alpha'\right)^{*l} a^l \middle/ l!\right)|l\rangle =$$

$$\exp\left(-\left(\left|\alpha'\right|^2 - \left|\alpha\right|^2\right)\middle/ 2\right)\exp\left(\beta a^+\right)D(\alpha)$$

$$\begin{pmatrix}|l\rangle + \left(\alpha-\alpha'\right)^* \sqrt{l}|l-1\rangle + \left(\alpha-\alpha'\right)^{*2}\sqrt{l(l-1)}|l-2\rangle \middle/ 2! + \\ \left(\alpha-\alpha'\right)^{*3}\sqrt{l(l-1)(l-2)}|l-3\rangle \middle/ 3! + ... + \left(\alpha-\alpha'\right)^{*l}\sqrt{l!}|0\rangle \middle/ l!\end{pmatrix} =$$

$$\exp\left(-\left(\left|\alpha'\right|^2 - \left|\alpha\right|^2\right)\middle/ 2\right)\exp\left(\beta a^+\right)$$

$$\begin{pmatrix}|l,\alpha\rangle + \left(\alpha-\alpha'\right)^* \sqrt{l}|l-1,\alpha\rangle + \left(\alpha-\alpha'\right)^{*2}\sqrt{l(l-1)}|l-2,\alpha.\rangle \middle/ 2! + \\ \left(\alpha-\alpha'\right)^{*3}\sqrt{l(l-1)(l-2)}|l-3,\alpha\rangle \middle/ 3! + ... + \left(\alpha-\alpha'\right)^{*l}\sqrt{l!}|0,\alpha\rangle \middle/ l!\end{pmatrix} \tag{9}$$

where $\alpha' = \alpha + \beta$ ($\beta = \alpha' - \alpha$) and $\alpha$, $\beta$ are the arbitrary numbers and $*$ is an operation of complex conjugation. Expression (9) consists of $l+1$ terms. Next step is to transform the state $\exp\left(\beta a^+\right)|l,\alpha\rangle$ through the displaced number states. It is possible to show the following identity

$$\exp\left(\beta a^+\right)|l,\alpha\rangle = \sum_{n=0}^{\infty}\frac{\left(\beta a^+\right)^n}{n!}|l,\alpha\rangle =$$

$$\exp\left(\beta\alpha^{\wedge}\right)\sum_{n=0}^{\infty}\frac{\beta^n}{\sqrt{n!}}\sqrt{\frac{(n+1)(n+2)...(n+l)}{l!}}|n+l,\alpha\rangle \tag{10}$$

holds. Inserting expression (10) into (9), one obtains final $\alpha-$representation of the displaced number state $|l,\alpha'\rangle$

$$|l,\alpha'\rangle = \exp\left(-\left(|\alpha'|^2-|\alpha|^2\right)\Big/2\right)\exp\left(\alpha'\alpha^*\right)$$

$$\left(\begin{array}{l}\sum_{n=0}^{\infty}\dfrac{\left(\alpha'-\alpha\right)^n}{\sqrt{n!}}\sqrt{\dfrac{(n+1)(n+2)...(n+l)}{l!}}|n+l,\alpha\rangle + \\[2mm] \left(\alpha-\alpha'\right)^*\sqrt{l}\sum_{n=0}^{\infty}\dfrac{\left(\alpha'-\alpha\right)^n}{\sqrt{n!}}\sqrt{\dfrac{(n+1)(n+2)...(n+l-1)}{(l-1)!}}|n+l-1,\alpha\rangle + \\[2mm] \left(\alpha-\alpha'\right)^{*2}\sqrt{l(l-1)}\Big/2!\sum_{n=0}^{\infty}\dfrac{\left(\alpha'-\alpha\right)^n}{\sqrt{n!}}\sqrt{\dfrac{(n+1)(n+2)...(n+l-2)}{(l-2)!}}|n+l-2,\alpha\rangle + \\[2mm] \left(\alpha-\alpha'\right)^{*3}\sqrt{l(l-1)(l-2)}\Big/3!\sum_{n=0}^{\infty}\dfrac{\left(\alpha'-\alpha\right)^n}{\sqrt{n!}}\sqrt{\dfrac{(n+1)(n+2)...(n+l-3)}{(l-3)!}}|n+l-3,\alpha\rangle + \\[2mm] ....+ \\[2mm] \left(\alpha-\alpha'\right)^{*(l-1)}\sqrt{l(l-1)(l-2)...2}\Big/(l-1)!\sum_{n=0}^{\infty}\dfrac{\left(\alpha'-\alpha\right)^n}{\sqrt{n!}}\sqrt{n+1}|n+1,\alpha\rangle + \\[2mm] \left(\alpha-\alpha'\right)^{*(l)}\sqrt{l!}\Big/l!\sum_{n=0}^{\infty}\dfrac{\left(\alpha'-\alpha\right)^n}{\sqrt{n!}}|n,\alpha\rangle\end{array}\right). \tag{11}$$

Consider more practical case of $\alpha'=0$. It follows from (11)

$$|l\rangle = |\Psi_1\rangle + |\Psi_2\rangle, \tag{12}$$

where a wave function $|\Psi_1\rangle$ is written

$$|\Psi_1\rangle = \exp\left(-|\alpha|^2\Big/2\right)\sum_{n=0}^{l-1}c_{\ln}|n,\alpha\rangle, \tag{13}$$

for the following values $0 \le n < l$. Wave amplitudes of the superposition (13) stems from (11)

$$c_{\ln} = \frac{\left(\alpha^*\right)^{l-n}}{\sqrt{l!}\sqrt{n!}}\sum_{k=0}^{n}(-1)^k C_n^k|\alpha|^{2k}\prod_{k}^{n-1}(l-n+k+1), \tag{14}$$

with $C_n^k = n!\Big/(k!(n-k)!)$ being binomial coefficients and symbol $\Pi$ means product of the numbers. Wave function $|\Psi_2\rangle$ is an infinite superposition of the displaced number states defined for $l \le n < \infty$

$$|\Psi_2\rangle = \exp\left(-|\alpha|^2\Big/2\right)\sum_{n=m}^{\infty}c_{\ln}|n,\alpha\rangle, \tag{15}$$

with the wave amplitudes

$$c_{\ln} = \frac{(-1)^{n-l}\alpha^{n-l}}{\sqrt{l!}\sqrt{n!}}\sum_{k=0}^{l}(-1)^k C_l^k|\alpha|^{2k}\prod_{k}^{l-1}(n-l+k+1). \tag{16}$$

Summing up formulas (12-16), we can finally construct unitary transformation matrix $U$ linking the base elements (6) with the number states

$$
\begin{Vmatrix} |0\rangle \\ |1\rangle \\ |2\rangle \\ |3\rangle \\ |4\rangle \\ |5\rangle \\ |6\rangle \\ ... \\ |n\rangle \\ ... \end{Vmatrix} = U \begin{Vmatrix} |0,\alpha\rangle \\ |1,\alpha\rangle \\ |2,\alpha\rangle \\ |3,\alpha\rangle \\ |4,\alpha\rangle \\ |5,\alpha\rangle \\ |6,\alpha\rangle \\ ... \\ |n,\alpha\rangle \\ ... \end{Vmatrix} = \exp\left(-|\alpha|^2/2\right) \begin{Vmatrix} c_{00} & c_{01} & c_{02} & c_{03} & c_{04} & c_{05} & c_{06} & ... & c_{0m} & ... \\ c_{10} & c_{11} & c_{12} & c_{13} & c_{14} & c_{15} & c_{16} & ... & c_{1m} & ... \\ c_{20} & c_{21} & c_{22} & c_{23} & c_{24} & c_{25} & c_{26} & ... & c_{2m} & ... \\ c_{30} & c_{31} & c_{32} & c_{33} & c_{34} & c_{35} & c_{36} & ... & c_{3m} & ... \\ c_{40} & c_{41} & c_{42} & c_{43} & c_{44} & c_{45} & c_{46} & ... & c_{4m} & ... \\ c_{50} & c_{51} & c_{52} & c_{53} & c_{54} & c_{55} & c_{56} & ... & c_{5m} & ... \\ c_{60} & c_{61} & c_{62} & c_{63} & c_{64} & c_{65} & c_{66} & ... & c_{6m} & ... \\ ... & ... & ... & ... & ... & ... & ... & ... & ... & \\ c_{n0} & c_{n1} & c_{n2} & c_{n3} & c_{n4} & c_{n5} & c_{n6} & ... & c_{nm} & ... \\ ... & ... & ... & ... & ... & ... & ... & ... & ... & \end{Vmatrix} \begin{Vmatrix} |0,\alpha\rangle \\ |1,\alpha\rangle \\ |2,\alpha\rangle \\ |3,\alpha\rangle \\ |4,\alpha\rangle \\ |5,\alpha\rangle \\ |6,\alpha\rangle \\ ... \\ |n,\alpha\rangle \\ ... \end{Vmatrix}.
$$
(17)

Matrix $U$ has its reverse $U^{-1}$ since the displacement operator (1) is unitary. The matrix account for reverse decomposition of the displaced number states over the number states

$$
\begin{Vmatrix} |0,\alpha\rangle \\ |1,\alpha\rangle \\ |2,\alpha\rangle \\ |3,\alpha\rangle \\ |4,\alpha\rangle \\ |5,\alpha\rangle \\ |6,\alpha\rangle \\ ... \\ |n,\alpha\rangle \\ ... \end{Vmatrix} = U^{-1} \begin{Vmatrix} |0\rangle \\ |1\rangle \\ |2\rangle \\ |3\rangle \\ |4\rangle \\ |5\rangle \\ |6\rangle \\ ... \\ |n\rangle \\ ... \end{Vmatrix} = \exp\left(-|\alpha|^2/2\right) \begin{Vmatrix} c_{00}^{*} & c_{10}^{*} & c_{20}^{*} & c_{30}^{*} & c_{40}^{*} & c_{50}^{*} & c_{60}^{*} & ... & c_{m0}^{*} & ... \\ c_{01}^{*} & c_{11}^{*} & c_{21}^{*} & c_{31}^{*} & c_{41}^{*} & c_{51}^{*} & c_{61}^{*} & ... & c_{m1}^{*} & ... \\ c_{02}^{*} & c_{12}^{*} & c_{22}^{*} & c_{32}^{*} & c_{42}^{*} & c_{52}^{*} & c_{62}^{*} & ... & c_{m2}^{*} & ... \\ c_{03}^{*} & c_{13}^{*} & c_{23}^{*} & c_{33}^{*} & c_{43}^{*} & c_{53}^{*} & c_{63}^{*} & ... & c_{m3}^{*} & ... \\ c_{04}^{*} & c_{14}^{*} & c_{24}^{*} & c_{34}^{*} & c_{44}^{*} & c_{54}^{*} & c_{64}^{*} & ... & c_{m4}^{*} & ... \\ c_{05}^{*} & c_{15}^{*} & c_{25}^{*} & c_{35}^{*} & c_{45}^{*} & c_{55}^{*} & c_{65}^{*} & ... & c_{m5}^{*} & ... \\ c_{06}^{*} & c_{16}^{*} & c_{26}^{*} & c_{36}^{*} & c_{46}^{*} & c_{56}^{*} & c_{66}^{*} & ... & c_{m6}^{*} & ... \\ ... & ... & ... & ... & ... & ... & ... & ... & ... & \\ c_{0n}^{*} & c_{1n}^{*} & c_{2n}^{*} & c_{3n}^{*} & c_{4n}^{*} & c_{5n}^{*} & c_{6n}^{*} & ... & c_{mn}^{*} & ... \\ ... & ... & ... & ... & ... & ... & ... & ... & ... & \end{Vmatrix} \begin{Vmatrix} |0\rangle \\ |1\rangle \\ |2\rangle \\ |3\rangle \\ |4\rangle \\ |5\rangle \\ |6\rangle \\ ... \\ |n\rangle \\ ... \end{Vmatrix}.
$$
(18)

Let us consider big values of the displacement amplitude $\alpha \gg 1$ to consider the states (2) with more energy (5) as macroscopic. Then decomposition $U^{-1}$ (18) can be fully recognized natural with classical point of view since it reflects that macroscopic state (object) is made up of microscopic states (stuff). But contrary idea that microscopic state (object) is composed of macroscopic states (with more energy) seems at least weird and downright bizarre with classical point of view. Nevertheless, it doesn't contradict quantum mechanics rules. Both direct (17) and reverse (18) transformations are legal in quantum mechanics.

Consider the partial elements of the matrix $U$ (17). So, matrix's elements of the first row

$$
c_{on}(\alpha) = \frac{(-1)^n \alpha^n}{\sqrt{n!}}
$$
(19)

are the coefficients of the Poisson distribution. They shape vacuum as infinite superposition of the displaced number states with amplitude $\alpha$. Elements of the second row of the matrix (17)

$$
c_{10}(\alpha) = \alpha^{*},
$$
(20)

$$
c_{1n}(\alpha) = \frac{(-1)^{n-1} \alpha^{n-1}}{\sqrt{n!}} \left(n - |\alpha|^2\right), \quad n \geq 1,
$$
(21)

determine decomposition of the single photon over the displaced number states with amplitude $\alpha$. Coefficients of $\alpha$ − representation of the two-photon state are given by

$$c_{20}(\alpha) = \frac{\alpha^{*2}}{\sqrt{2!}}, \tag{22}$$

$$c_{21}(\alpha) = \frac{\alpha^{*}}{\sqrt{2!}}\left(2 - |\alpha|^2\right), \tag{23}$$

$$c_{2n}(\alpha) = \frac{(-1)^{m-2}\alpha^{m-2}}{\sqrt{2!}\sqrt{n!}}\left(n(n-1) - 2n|\alpha|^2 + |\alpha|^4\right), \quad n \geq 2. \tag{24}$$

Finally, consider wave amplitudes of three photon state $|3\rangle$

$$c_{30}(\alpha) = \frac{\alpha^{*3}}{\sqrt{3!}}, \tag{25}$$

$$c_{31}(\alpha) = \frac{\alpha^{*2}}{\sqrt{3!}}\left(3 - |\alpha|^2\right), \tag{26}$$

$$c_{32}(\alpha) = \frac{\alpha^{*}}{\sqrt{3!}\sqrt{2!}}\left(6 - 6|\alpha|^2 + |\alpha|^4\right), \tag{27}$$

$$c_{3n}(\alpha) = \frac{(-1)^{m-3}\alpha^{m-3}}{\sqrt{3!}\sqrt{n!}}\left(n(n-1)(n-2) - 3n(n-1)|\alpha|^2 + 3n|\alpha|^4 - |\alpha|^6\right), \quad n \geq 3. \tag{28}$$

Comparing formulas (19, 20, 22, 25), we can recognize the coefficients of the Poisson distribution of the coherent state (first row of the reverse matrix $U^{-1}$ (18)).

Probability to observe $n$ photons displaced on $\alpha$ in $l$ − photon state is defined by the corresponding wave amplitude $c_{\ln}$ modulo squared

$$P_{\ln}(\alpha) = |c_{\ln}(\alpha)|^2. \tag{29}$$

Normalization condition $\langle l | l \rangle = 1$ for arbitrary $l$ − photon state

$$\sum_{n=0}^{\infty} P_{\ln}(\alpha) = 1, \tag{30}$$

is implemented. Figures 1(a-f) and 2 (a-f) reveal probability distributions of single and three-photon states over the displaced number states for different displacement amplitudes. The distributions manifest typical inherent features especially with increase of the displacement amplitude, when $l + 1$ peaks are beheld in $l$ − photon state.

## 2.2 $\alpha$ − representation of two-mode squeezed state

Decompositions (17, 18) with elements (14, 16) can become base for construction of $\alpha$ − representation of other states. But derivation of the expressions is not trivial. Consider $\alpha$ − representation of TMSV [23] defined through two-mode squeezed operator

$$|\Psi\rangle_{12} = S_{12}(r)|00\rangle_{12} = (1/\cosh r)\sum_{n=0}^{\infty}(\tanh r)^n |n\rangle_1 |n\rangle_2 = (1/\cosh r)A^T B, \tag{31}$$

where operator $S_{12}(r)$ is given by [23]

$$S_{12}(r) = \exp\left(r\left(a_1^+ a_2^+ - a_1 a_2\right)\right) \tag{32}$$

with $r$ being the squeezing parameter. The expression (31) can be rewritten in matrix form, where infinite vector-columns are the following

$$A^T = \left| \begin{array}{ccccc} |0\rangle & \tanh r|1\rangle & (\tanh r)^2|2\rangle & ... & (\tanh r)^l|l\rangle & ... \end{array} \right|, \tag{33}$$

$$B = \left| \ |0\rangle \quad |1\rangle \quad |2\rangle \quad ... \quad |n\rangle \quad ... \right|^T, \tag{34}$$

where symbol $T$ means matrix transposition.

Using the matrix $U$ (17), column $B$ can be presented as product of the transformation matrix $U$ (17) on column of the displaced number states. It enables to rewrite $A^T B$ as

$$A^T B = \exp\left(-|\alpha|^2/2\right) \left| \ |0\rangle \quad \tanh r|1\rangle \quad \left(\tanh r\right)^2|2\rangle \quad ... \quad \left(\tanh r\right)^l|l\rangle \quad ... \right|$$

$$\begin{vmatrix} c_{00} & c_{01} & c_{02} & ... & c_{0l} & ... \\ c_{10} & c_{11} & c_{12} & ... & c_{1l} & ... \\ c_{20} & c_{21} & c_{22} & ... & c_{2l} & ... \\ ... & ... & ... & ... & ... & ... \\ c_{l0} & c_{l1} & c_{l2} & ... & c_{ll} & ... \\ ... & ... & ... & ... & ... & ... \end{vmatrix} \begin{vmatrix} |0,\alpha\rangle \\ |1,\alpha\rangle \\ |2,\alpha\rangle \\ ... \\ |l,\alpha\rangle \\ ... \end{vmatrix}. \tag{35}$$

Let us introduce the following infinite column of the states

$$C^T = \left| \ |\psi_0\rangle \quad |\psi_1\rangle \quad |\psi_2\rangle \quad ... \quad |\psi_n\rangle \quad ... \right| =$$

$$\left| \ |0\rangle \quad \tanh r|1\rangle \quad \left(\tanh r\right)^2|2\rangle \quad ... \quad \left(\tanh r\right)^l|l\rangle \quad ... \right| \begin{vmatrix} c_{00} & c_{01} & c_{02} & ... & c_{0l} & ... \\ c_{10} & c_{11} & c_{12} & ... & c_{1l} & ... \\ c_{20} & c_{21} & c_{22} & ... & c_{2l} & ... \\ ... & ... & ... & ... & ... & ... \\ c_{l0} & c_{l1} & c_{l2} & ... & c_{ll} & ... \\ ... & ... & ... & ... & ... & ... \end{vmatrix}, \tag{36}$$

to finally present TMSV (31) in the form

$$|\Psi\rangle_{12} = \left(\exp\left(-|\alpha|^2/2\right)/\cosh r\right) C^T B = \left(\exp\left(-|\alpha|^2/2\right)/\cosh r\right)\sum_{n=0}^{\infty}|\psi_n\rangle_1|n,\alpha\rangle_2, \tag{37}$$

where the wave function $|\psi_n\rangle$ is infinite superposition of the number states

$$|\psi_n\rangle = \sum_{l=0}^{\infty} c_{\ln}(\alpha)\left(\tanh r\right)^l|l\rangle. \tag{38}$$

Representation (37) may be called $\alpha-$ representation of the TMSV but it is hardly comfortable for practical applications. We have to manipulate with amplitudes of the state (38) to obtain more convenient form of the $\alpha-$ representation of the TMSV.

The wave amplitudes of the superposition (38) are formed from elements of $n$ column of the matrix (17) multiplied by factors $\left(\tanh r\right)^l$. Contribution of the additional factor occurs from squeezing operator (32). It is possible to exhibit that the wave amplitude $c_{\ln}(\alpha)\left(\tanh r\right)^l$ for arbitrary value of the parameter $l$ can be decomposed into finite series of the basic functions

$$\left\{c_{l0}^*(\beta), c_{l1}^*(\beta), c_{l2}^*(\beta), c_{l3}^*(\beta), ..., c_{\ln}^*(\beta)\right\}, \tag{39}$$

where $\beta = \alpha^* \tanh r$. Set of the base functions (39) consists of $n+1$ terms. Mathematical proof of the fact is not trivial and is not presented here. The argument of the base functions (39) is a product of $\alpha^*$ on $\tanh r$ $\left(\alpha^* \tanh r\right)$, while amplitudes $c_{\ln}(\alpha)$ are the functions of only $\alpha$. Amplitudes $a_{\ln}(\alpha)$ of decomposition over the base functions (39) are calculated from matrix equations that are not presented here for their complexity. Finally, the matrix elements $c_{\ln}(\alpha)$ are decomposed as

$$c_{\text{ln}}(\alpha)(\tanh r)^l = \sum_{k=0}^{n} a_{nk} c_{lk}^*(\beta), \tag{40}$$

where the amplitudes of the decomposition are the following

$$a_{n0}(\alpha) = \alpha^n \left((\tanh r)^2 - 1\right)^n / \sqrt{n!}, \tag{41}$$

$$a_{n1}(\alpha) = n\alpha^{n-1} \tanh r \left((\tanh r)^2 - 1\right)^{n-1} / \sqrt{n!}, \tag{42}$$

$$a_{n2}(\alpha) = n(n-1)\alpha^{n-2}(\tanh r)^2 \left((\tanh r)^2 - 1\right)^{n-2} / \left(\sqrt{2!}\sqrt{n!}\right), \tag{43}$$

$$a_{n3}(\alpha) = n(n-1)(n-2)\alpha^{n-3}(\tanh r)^3 \left((\tanh r)^2 - 1\right)^{n-3} / \left(\sqrt{3!}\sqrt{n!}\right), \tag{44}$$

$$a_{nk}(\alpha) = n(n-1)(n-2)...(n-k)\alpha^{n-k}(\tanh r)^k \left((\tanh r)^2 - 1\right)^{n-k} / \left(\sqrt{k!}\sqrt{n!}\right), \tag{45}$$

$$a_{nn}(\alpha) = (\tanh r)^n. \tag{46}$$

Make use of the following symbolic matrix

$$\begin{array}{c|cccccccc|c}
 & a_{n0} & a_{n1} & a_{n2} & a_{n3} & a_{n4} & ... & a_{nn} & & |\psi_n\rangle \\
\hline
|0\rangle & c_{00}^*(\beta) & c_{01}^*(\beta) & c_{02}^*(\beta) & c_{03}^*(\beta) & c_{04}^*(\beta) & ... & c_{0n}^*(\beta) & & c_{0n}(\alpha) \\
|1\rangle & c_{10}^*(\beta) & c_{11}^*(\beta) & c_{12}^*(\beta) & c_{13}^*(\beta) & c_{14}^*(\beta) & ... & c_{1n}^*(\beta) & & c_{1n}(\alpha)\tanh r \\
|2\rangle & c_{20}^*(\beta) & c_{21}^*(\beta) & c_{22}^*(\beta) & c_{23}^*(\beta) & c_{24}^*(\beta) & ... & c_{2n}^*(\beta) & & c_{2n}(\alpha)(\tanh r)^2 \\
|3\rangle & c_{30}^*(\beta) & c_{31}^*(\beta) & c_{32}^*(\beta) & c_{33}^*(\beta) & c_{34}^*(\beta) & ... & c_{3n}^*(\beta) & & c_{3n}(\alpha)(\tanh r)^3 \\
|4\rangle & c_{40}^*(\beta) & c_{41}^*(\beta) & c_{42}^*(\beta) & c_{43}^*(\beta) & c_{44}^*(\beta) & ... & c_{4n}^*(\beta) & & c_{4n}(\alpha)(\tanh r)^4 \\
|5\rangle & c_{50}^*(\beta) & c_{51}^*(\beta) & c_{52}^*(\beta) & c_{53}^*(\beta) & c_{54}^*(\beta) & ... & c_{5n}^*(\beta) & & c_{5n}(\alpha)(\tanh r)^5 \\
... & ... & ... & ... & ... & ... & & ... & & ... \\
|k\rangle & c_{k0}^*(\beta) & c_{k1}^*(\beta) & c_{k2}^*(\beta) & c_{k3}^*(\beta) & c_{k4}^*(\beta) & ... & c_{kn}^*(\beta) & & c_{kn}(\alpha)(\tanh r)^k \\
... & ... & ... & ... & ... & ... & & ... & & ... \\
|n\rangle & c_{n0}^*(\beta) & c_{n1}^*(\beta) & c_{n2}^*(\beta) & c_{n3}^*(\beta) & c_{n4}^*(\beta) & ... & c_{nm}^*(\beta) & & c_{nn}(\alpha)(\tanh r)^n \\
... & ... & ... & ... & ... & ... & & ... & & ... \\
|l\rangle & c_{l0}^*(\beta) & c_{l1}^*(\beta) & c_{l2}^*(\beta) & c_{l3}^*(\beta) & c_{l4}^*(\beta) & ... & c_{ln}^*(\beta) & & c_{ln}(\alpha)(\tanh r)^l \\
 & |0,\beta\rangle & |1,\beta\rangle & |2,\beta\rangle & |3,\beta\rangle & |4,\beta\rangle & ... & |n,\beta\rangle & & |\psi_n\rangle
\end{array} \tag{47}$$

to elucidate details of the transformation with the states. The matrix (47) consists of finite number $n+3$ of columns and infinite number of the rows. First column of the matrix (47) is a column of number states from 0 up to $\infty$. Last is a column with elements $c_{\text{ln}}(\alpha)(\tanh r)^l$, where $l$ is varied from 0 up to $\infty$. It entirely describes the state $|\psi_n\rangle$ (38) in terms of the number states. First row involves amplitudes of the decomposition (41-46) over the base functions (39) and also the state (38). Next rows except for elements of the first and last columns are constructed from the corresponding base functions (39). Comparing the column elements under corresponding amplitude $a_{nm}$ with $m$ varying from 0 up to $n$ and rows of the reverse matrix (18), we can notice they are identical. If we multiply the column on the factor $\exp\left(-|\beta|^2/2\right)$, we write down the final row from the displaced number states. Thus, the state (38) can be rewritten as finite $n+1$ superposition of the displaced number states

$$|\psi_n\rangle = \exp\left(|\beta|^2/2\right) a_{n0} \sum_{k=0}^{n} b_{nk} |k,\beta\rangle = \exp\left(|\beta|^2/2\right) a_{n0} D(\beta) \sum_{k=0}^{n} b_{nk} |k\rangle, \tag{48}$$

where

$$b_{n0}(\alpha) = 1, \tag{49}$$

$$b_{n1}(\alpha) = a_{n1}(\alpha)/a_{n0}(\alpha) = n\gamma , \tag{50}$$

$$b_{n2}(\alpha) = a_{n2}(\alpha)/a_{n0}(\alpha) = n(n-1)\gamma^2/\sqrt{2!} , \tag{51}$$

$$b_{n3}(\alpha) = a_{n3}(\alpha)/a_{n0}(\alpha) = n(n-1)(n-2)\gamma^3/\sqrt{3!}, \tag{52}$$

$$b_{nk}(\alpha) = a_{nk}(\alpha)/a_{n0}(\alpha) = n(n-1)(n-2)...(n-k+1)\gamma^k/\sqrt{k!}, \tag{53}$$

$$b_{nn}(\alpha) = a_{nn}(\alpha)/a_{n0}(\alpha) = \sqrt{n!}\gamma^n , \tag{54}$$

and

$$\gamma = \tanh r/\left(\alpha\left((\tanh r)^2 - 1\right)\right). \tag{55}$$

Consider the finite sum of the number states

$$\sum_{k=0}^{n} b_{nk}|k\rangle . \tag{56}$$

We can recognize it is the operator $\left(1 + \gamma a^+\right)$ in power $n$ acting on vacuum state

$$\sum_{k=0}^{n} b_{nk}|k\rangle = \left(1 + \gamma a^+\right)^n|0\rangle . \tag{57}$$

The following identity

$$a_{n0}\gamma^n = (\tanh r)^n/\sqrt{n!} \tag{58}$$

takes place. It gives a possibility to finally rewrite the state $|\psi_n\rangle$ (38) as

$$|\psi_n\rangle = \exp\left(|\beta|^2/2\right)D(\beta)(\tanh r)^n\left(a^+ - \delta^*\right)^n|0\rangle/\sqrt{n!}, \tag{59}$$

where the parameter $\delta$ is determined by

$$\delta = \alpha^*\left(1 - (\tanh r)^2\right)/\tanh r . \tag{60}$$

Finally, $\alpha-$representation of the two-mode squeezed state (31) can be presented as

$$|\Psi\rangle_{12} = \left(\exp\left(-(\sinh r)^2|\delta|^2/2\right)/\cosh r\right)D_1\left(\alpha^*\tanh r\right)D_2(\alpha)$$

$$\sum_{n=0}^{\infty}\left((\tanh r)^n\left(a_1^+ - \delta^*\right)^n|0\rangle_1/\sqrt{n!}\right)|n\rangle_2 = \tag{61}$$

$$\left(\exp\left(-(\sinh r)^2|\delta|^2/2\right)/\cosh r\right)D_1\left(\alpha^*\tanh r\right)D_2(\alpha)\sum_{n=0}^{\infty}(\tanh r)^n N_n|\Psi_n\rangle|n\rangle_2$$

where normalized state $|\Psi_n\rangle$ has a form

$$|\Psi_n\rangle = A^{+n}|0\rangle/\left(N_n\sqrt{n!}\right) = \left(1/N_n\right)$$

$$\left(|n\rangle + \sum_{l=1}^{n}\left((-1)^l\delta^{*ll}\sqrt{n(n-1)(n-2)...(n-l+1)}/l!\right)|n-l\rangle\right) \tag{62}$$

and normalization factor is given by

$$N_n = \left(1 + \sum_{l=1}^{n}\frac{|\delta|^{2l}n(n-1)(n-2)...(n-l+1)}{|l|^2}\right)^{1/2} . \tag{63}$$

Now, expression (61) can be named $\alpha-$representation of the TMSV (31). The representation is dependent on parameters $\delta$ and $r$. The displacement amplitudes $\alpha^*\tanh r$ and $\alpha$ in neighboring correlated modes of the TMSV differ from each other. We can swap the displacement amplitudes in the modes due to symmetry of the state (31). Consider the case of $\alpha = 0$ that gives $\delta = 0$. Then, the state (61) acquires its initial form (31) and can be called $0-$representation of the TMSV. Parameter $\delta$ approaches to zero $\delta \to 0$ when $r \to \infty$ to result in

$$|\Psi\rangle_{12} = (1/\cosh r) D_1(\alpha^*) D_2(\alpha) \sum_{n=0}^{\infty} |n\rangle_1 |n\rangle_2 . \tag{64}$$

Limit case (64) corresponds to the original (perfectly correlated and maximally entangled, but unphysical) EPR state [23] in $\alpha$ – representation.

Amplitudes of the TMSV in $\alpha$ – representation are given by

$$p_n(\delta, r) = \left( \exp\left( -(\sinh r)^2 |\delta|^2 /2 \right) / \cosh r \right) (\tanh r) N_n(\delta, r). \tag{65}$$

They are even functions of the parameter $\alpha$

$$p_n(\delta, r) = p_n(-\delta, r). \tag{66}$$

Probability to look for correlated displaced number states $D(\alpha^* \tanh r) D_2(\alpha) |n\rangle_1 |n\rangle_2$ in TMSV is given by

$$P_n(\delta, r) = (\tanh r)^{2n} |N_n|^2 \exp\left( -(\sinh r)^2 |\delta|^2 \right) / (\cosh r)^2 . \tag{67}$$

It is possible to show normalization condition

$$\sum_{n=0}^{\infty} P_n(\delta, r) = 1 \tag{68}$$

is accomplished for any values of $\delta$ and $r$. Probability distributions $P_n(\delta, r)$ in dependency on $n$ for different values of $\delta$ and $r$ are shown in figure 3(a-f). The probability distribution $P_n(\delta, r)$ has maximum value for some value of $n$. Maximal value of $P_n(\delta, r)$ depends on $\delta$ and $r$ in complex fashion.

## 2.3 $\alpha$ – representations of superposition of vacuum and single photon

Consider the following balanced superpositions of vacuum and single photon

$$|\Delta_{\pm}\rangle = \left( |0\rangle \pm |1\rangle \right) / \sqrt{2} . \tag{69}$$

The states (69) are orthogonal to each other. If we make use of first and second rows of the matrix (17) whose elements are given by the formulas (19-21), we can obtain their $\alpha$ – representations as

$$|\Delta_+\rangle = \frac{\exp\left( -|\alpha|^2 /2 \right)}{\sqrt{2}} \left( \begin{array}{l} \left(1 + \alpha^*\right) |0, \alpha\rangle + \\ \sum_{n=1}^{\infty} \frac{(-1)^n \alpha^{n-1}}{\sqrt{n!}} \left( \alpha - \left(n - |\alpha|^2\right) \right) |n, \alpha\rangle \end{array} \right), \tag{70}$$

$$|\Delta_-\rangle = \frac{\exp\left( -|\alpha|^2 /2 \right)}{\sqrt{2}} \left( \begin{array}{l} \left(1 - \alpha^*\right) |0, \alpha\rangle + \\ \sum_{n=1}^{\infty} \frac{(-1)^n \alpha^{n-1}}{\sqrt{n!}} \left( \alpha + \left(n - |\alpha|^2\right) \right) |n, \alpha\rangle \end{array} \right). \tag{71}$$

Amplitudes of the superpositions are the following

$$f_0^{(+)}(\alpha) = \exp\left( -|\alpha|^2 /2 \right) \left(1 + \alpha^*\right) / \sqrt{2} , \tag{72}$$

$$f_n^{(+)}(\alpha) = \exp\left( -|\alpha|^2 /2 \right) \left( (-1)^n \alpha^{n-1} \left( \alpha - \left(n - |\alpha|^2\right) \right) \right) / \left( \sqrt{2} \sqrt{n!} \right), \tag{73}$$

for the state (70) and

$$f_0^{(-)}(\alpha) = \exp\left( -|\alpha|^2 /2 \right) \left(1 - \alpha^*\right) / \sqrt{2} , \tag{74}$$

$$f_n^{(-)}(\alpha) = \exp\left( -|\alpha|^2 /2 \right) \left( (-1)^n \alpha^{n-1} \left( \alpha + \left(n - |\alpha|^2\right) \right) \right) / \left( \sqrt{2} \sqrt{n!} \right), \tag{75}$$

for the state (71). Here, subscript is its number and superscript concerns sign in superpositions (70, 71). They satisfy the condition

$$f_{2n}^{(+)}(\alpha) = f_{2n}^{(-)}(-\alpha), \tag{76}$$

$$f_{2n+1}^{(+)}(\alpha) = -f_{2n+1}^{(-)}(-\alpha), \tag{77}$$

that means the even amplitudes are even and odd amplitudes are odd by change of the displacement amplitude.

Expressions (70, 71) are converted to (69) in the case of $\alpha = 0$. This implies the state (69) can be named $0-$ representation of the superposition of vacuum and single photon. Probability distribution over $n-$ number states displaced on $\alpha$ is defined by

$$P_{n\pm}(\alpha) = \frac{\exp\left(-|\alpha|^2\right)}{2} \frac{|\alpha|^{2(n-1)}}{n!} \left|\alpha \mp \left(n - |\alpha|^2\right)\right|^2, \tag{78}$$

which is even function with regard to sign change of $\alpha$

$$P_{n+}(\alpha) = P_{n-}(-\alpha) \tag{79}$$

Normalization condition leads to

$$\sum_{n=0}^{\infty} P_{n\pm}(\alpha) = 1. \tag{80}$$

Plots of $P_{n+}(\alpha)$ and $P_{n-}(\alpha)$ are presented in figures 4(a-f) and 5(a-f), respectively, for different values of $\alpha$. Probabilities of the displaced vacuum and displaced single photon prevail over other probabilities in the case of small value of $\alpha$. The probability distributions manifest new properties with $\alpha$ growing. So, two peaks are shifted to side of larger values of $n$ and one of the peaks is higher of other. Relative location of the peaks is interchanged by places in dependency on type of the state.

## 3. Realization of elementary gates with help of $\alpha-$ representation of the states

Above, derivation of $\alpha-$ representation of known states is presented. Let us discuss physical meaning of the new representation. The physical information is contained in wave amplitudes of initial state. All what we wish to know about physical system in quantum mechanics is in column vector of amplitudes of the state of the system. Vector columns of amplitudes in different bases related to each other by unitary operator are connected with corresponding matrix. For example, transposed matrix (17) is used to get column vector of amplitudes of the state in basis of the displaced number states. This means there are different ways of encoding the same information enclosed in the state. It may resemble $x-$ and $p-$ representations of the wave function of the quantum particle.

All we need to work with amplitudes in different bases is set of observables, in particular, projection operators on displaced number states $M_n = |n, \alpha\rangle\langle n, \alpha|$. The operators can be realized in optics by means of mixing of initial state with coherent state of large amplitude [9] on highly transmitting beam splitter followed by avalanche photodiodes (APD). APD possesses high quantum efficiency but can only discriminate the presence of radiation from the vacuum. It can be used to reconstruct the photon statistics but cannot be used as photon counters. Number state $|n\rangle$ can be measured by photon number resolving detector (PNRD) able to distinguish outcomes from different number states. Now, PNRD are only prototype with restricted possibilities [28]. System of unbalanced beam splitter (UBS) and balanced beam splitters (BBSs) followed by APDs may be partly used instead of PNRD to discriminate the number states [20, 21].

Operator $A^{+n}$ (4) with amplitude $\delta$ affecting vacuum in unmeasured mode of a composite system (61) is generated provided that displaced number state $|n, \alpha\rangle$ is measured in the neighboring mode. Efficiency of the $\alpha-$ representation is not restricted with the example. Symmetry properties of the amplitudes regarding sign change of the displacement amplitude in $\alpha-$ representation can be useful for generation of new superposed states. Indeed, superposition of coherent states with equal modulo but opposite on sign amplitudes can simultaneously displace target state on $\pm\alpha$ after interaction with it on beam splitter. Output state becomes hybrid. It is formed from the base elements of different two-dimensional Hilbert spaces.

## 3.1 Realization of two-qubit control-sign gate

Consider an optical scheme in figure 6. The scheme is a Mach-Zehnder interferometer in arms of which control qubit additionally interacts with TMSV through the beam splitters. TMSV occupies modes 3 and 4. The similar scheme with second-order crystals in arms of the interferometer was used in [29] in order to conditionally generate maximally entangled state of two photons provided that pumping photon is registered. Input qubit is in mode 1, while mode 2 is launched to the interferometer in vacuum state. Initial qubit splits on input beam splitter $B_{12}^{'}$ and travels simultaneously along both interferometer's modes to the output beam splitter $B_{12}$ to combine on it. Depending on the beam splitter parameters, in general case, output qubit in both modes goes out from the interferometer. But if the input balanced splitter is chosen so that $B_{12}^{'} = B_{12}^{-1}$, then output qubit certainly comes out from input mode.

Let us consider the UBSs with the following matrixes

$$B_{12}^{'} = \begin{vmatrix} t_1 & r_1 \\ -r_1 & t_1 \end{vmatrix}, \tag{81}$$

$$B_{12} = \begin{vmatrix} t_1 & -r_1 \\ r_1 & t_1 \end{vmatrix}, \tag{82}$$

with amplitudes of transparency and reflectivity

$$t_1 = \tanh s / \sqrt{1 + (\tanh s)^2}, \tag{83}$$

$$r_1 = \tanh s / \sqrt{1 + (\tanh s)^2}, \tag{84}$$

respectively, where $s$ is a squeezing parameter of the TMSV. Two UBSs $B_{13}$ and $B_{24}$ are inserted inside the interferometer to organize interaction of the input state with TMSV

$$B_{13} = \begin{vmatrix} t & -r\exp(-i\varphi) \\ r\exp(i\varphi) & t \end{vmatrix}, \tag{85}$$

$$B_{24} = \begin{vmatrix} t & -r\exp(i\varphi) \\ r\exp(-i\varphi) & t \end{vmatrix}, \tag{86}$$

where $t \to 1$ is an amplitude of transparency, $r \to 0$ is an amplitude of reflectivity and $\varphi$ is an additional phase shift.

The following coherent states

$$\left| \overset{\wedge}{0} \right\rangle = \left| 0, \alpha t \sqrt{1 + (\tanh s)^2} / r \right\rangle, \tag{87}$$

$$\left| \overset{\wedge}{1} \right\rangle = \left| 0, -\alpha t \sqrt{1 + (\tanh s)^2} / r \right\rangle, \tag{88}$$

are chosen as the base elements of input two-dimensional Hilbert space , $(\alpha > 0)$. The input base states are asymptotically orthogonal since their overlapping $\left| \left\langle \overset{\wedge}{0} | \overset{\wedge}{1} \right\rangle \right|^2 = \exp\left( -4\alpha t \sqrt{1 + (\tanh s)^2} / r \right) \to 0$ exponentially drops with argument growing. The

condition $r \to 0$ guaranties almost perfect orthogonality of the coherent states (87, 88). Therefore with practical point of view, we can consider the coherent states (87, 88) orthogonal. Superposition of the base elements can be written as

$$\left|\Psi_1\right\rangle = a\left|\hat{0}\right\rangle + b\left|\hat{1}\right\rangle =$$
$$\left|\Psi_{ab}\left(\alpha t\sqrt{1+\left(\tanh s\right)^2}\middle/r\right)\right\rangle = a\left|0,\alpha t\sqrt{1+\left(\tanh s\right)^2}\middle/r\right\rangle + b\left|0,-\alpha t\sqrt{1+\left(\tanh s\right)^2}\middle/r\right\rangle,$$

(89)

where normalization condition $\left|a\right|^2 + \left|b\right|^2 = 1$ holds. Another two-dimensional Hilbert space is formed from vacuum and single photon

$$\left|\hat{0}_1\right\rangle = \left|0\right\rangle,$$

(90)

$$\left|\hat{1}_1\right\rangle = \left|1\right\rangle,$$

(91)

being the base elements. Balanced superpositions of the basic states (90, 91) are just the states (69) and arbitrary superposition is defined by

$$\left|\Psi_2\right\rangle = a_1\left|\hat{0}_1\right\rangle + b_1\left|\hat{1}_1\right\rangle = a_1\left|0\right\rangle + b_1\left|1\right\rangle,$$

(92)

with $\left|a_1\right|^2 + \left|b_1\right|^2 = 1$.

The coherent superposition (89) is a control qubit in figure 6. Target qubit (92) is initially part of the TMSV (61). Indeed, consider the term in (61) with $n = 1$

$$D_3\left(\alpha^*\tanh s\right)D_4\left(\alpha\right)\left(a^+ - \delta^*\right)\left|0\right\rangle_3\left|1\right\rangle_4.$$

(93)

To pull out the qubit (92) from TMSV, two displacement operators $D_3\left(-\alpha^*\tanh s\right)D_4\left(-\alpha\right)$ are applied followed by projective measurement on single photon in mode 4. Given that outcome registered, the output state after the measurement is the qubit (92) with $a_1^+ = -\delta^*\middle/\sqrt{1+\left|\delta\right|^2}$ and $b_1^+ = 1\middle/\sqrt{1+\left|\delta\right|^2}$. Change sign of the displacement amplitude on opposite $\alpha \to -\alpha$ to deal with the case

$$D_3\left(-\alpha^*\tanh s\right)D_4\left(-\alpha\right)\left(a^+ + \delta^*\right)\left|0\right\rangle_3\left|1\right\rangle_4.$$

(94)

Now, we follow the same procedure with operators $D_3\left(\alpha^*\tanh s\right)D_4\left(\alpha\right)$ to generate qubit (92) with $a_1^- = -a_1^- = \delta^*\middle/\sqrt{1+\left|\delta\right|^2}$ and $b_1^- = b_1^- = 1\middle/\sqrt{1+\left|\delta\right|^2}$. Only amplitude of the vacuum state is varied on opposite with change of the displacement amplitude $\alpha \to -\alpha$. Sets of the displacement operators $D_3\left(\mp\alpha^*\tanh s\right)D_4\left(\mp\alpha\right)$ are produced with help of the coherent states (87, 88) combined with target ones on highly transmissive beam splitters, where relative phase of the qubit (92) is assigned by the phase shift of the UBSs (85, 86). Final registration of single photon gives a birth to constructive interference. Last UBS of the Mach-Zehnder interferometer in figure 6 joins coherent states into one mode. Additional phase shifter on $\pi$ $P(\pi) = \exp\left(ia^+a\right)$ in coherent mode is used to change sign $\left|0,\alpha\right\rangle \leftrightarrow \left|0,-\alpha\right\rangle$. Finally, we can write the following chain of transformations

$$P_1(\pi)M_4^{(1)}B_{12}B_{13}B_{24}B_{12}^-\left(\left|\Psi_{ab}\left(\alpha t\sqrt{1+\left(\tanh s\right)^2}\middle/r\right)\right\rangle_1\left|0\right\rangle_2 S(r)\left|00\right\rangle_{34}\right) \to$$
$$a\left|0,-\gamma\right\rangle_1\left(a_1\left|0\right\rangle_2 + b_1\left|1\right\rangle_1\right) + b\left|0,-\gamma\right\rangle_1\left(-a_1\left|0\right\rangle_2 + b_1\left|1\right\rangle_2\right) =$$
$$aa_1\left|0,-\gamma\right\rangle_1\left|0\right\rangle_2 + ab_1\left|0,-\gamma\right\rangle_1\left|1\right\rangle_2 - ba_1\left|0,-\gamma\right\rangle_1\left|0\right\rangle_2 + bb_1\left|0,-\gamma\right\rangle_1\left|1\right\rangle_2,$$

(95)

where $M_4^{(1)} = \left(|1\rangle\langle 1|\right)_4$ is a projection operator on single photon state in mode 4 and $\gamma = \alpha\sqrt{1+(\tanh s)^2}\big/r$ is an amplitude of the output coherent state that differs initial one (87, 88) by factor $t \approx 1$. The final state (95) is the output state of control-sign gate fir control and target qubits provided that permutable target basis $\left|\hat{0}_1\right\rangle \leftrightarrow \left|\hat{1}_1\right\rangle$ (90, 91) is used. Then, the expression (95) can be rewritten as

$$CZ\left(|\Psi_1\rangle|\Psi_2\rangle\right) = CZ\left(a\left|\hat{0}\right\rangle + b\left|\hat{1}\right\rangle\right)_1\left(a_1\left|\hat{0}_1\right\rangle + b_1\left|\hat{1}_1\right\rangle\right)_2 =$$
$$aa_1\left|\hat{0}\right\rangle_1\left|\hat{0}\right\rangle_1 + ab_1\left|\hat{0}\right\rangle_1\left|\hat{1}\right\rangle_1 + ba_1\left|\hat{1}\right\rangle_1\left|\hat{0}\right\rangle_1 - bb_1\left|\hat{1}\right\rangle_1\left|\hat{1}\right\rangle_1$$

(96a)

or in matrix representation

$$CZ\begin{vmatrix} aa_1 \\ ab_1 \\ ba_1 \\ bb_1 \end{vmatrix} = \begin{vmatrix} 1 & 0 & 0 & 0 \\ 0 & 1 & 0 & 0 \\ 0 & 0 & 1 & 0 \\ 0 & 0 & 0 & -1 \end{vmatrix}\begin{vmatrix} aa_1 \\ ab_1 \\ ba_1 \\ bb_1 \end{vmatrix} = \begin{vmatrix} aa_1 \\ ab_1 \\ ba_1 \\ -bb_1 \end{vmatrix},$$

(96b)

where the control qubit $|\Psi_1\rangle$ can be considered as macroscopic and the target qubit $|\Psi_2\rangle$ is microscopic. The operation (96) is successfully realized provided that the displaced single photon is subtracted from TMSV.

## 3.2 Realization of Hadamard matrix with hybrid states

Consider the partial case of $a = b = 1/\sqrt{2}$ $\left(a = -b = 1/\sqrt{2}\right)$, $\delta = \pm 1$ and as consequence $a_1 = b_1 = 1/\sqrt{2}$ in equation (95). Then, the balanced hybrid states are generated

$$|\Phi_+\rangle_{12} = \left(\begin{array}{l}\left|0, \alpha\sqrt{1+(\tanh s)^2}\big/r\right\rangle_1\left(|0\rangle + |1\rangle\right)_2 + \\ \left|0, -\alpha\sqrt{1+(\tanh s)^2}\big/r\right\rangle_1\left(-|0\rangle + |1\rangle\right)_2\end{array}\right)\bigg/2 =$$
$$\left(\begin{array}{l}\left(\left|0, \alpha\sqrt{1+(\tanh s)^2}\big/r\right\rangle_1 - \left|0, -\alpha\sqrt{1+(\tanh s)^2}\big/r\right\rangle_1\right)|0\rangle_2 + \\ \left(\left|0, \alpha\sqrt{1+(\tanh s)^2}\big/r\right\rangle_2 + \left|0, -\alpha\sqrt{1+(\tanh s)^2}\big/r\right\rangle_1\right)|1\rangle_2 + \end{array}\right)\bigg/2,$$

(97)

$$|\Phi_-\rangle_{12} = \left(\begin{array}{l}\left|0, \alpha\sqrt{1+(\tanh s)^2}\big/r\right\rangle_1\left(|0\rangle + |1\rangle\right)_2 - \\ \left|0, -\alpha\sqrt{1+(\tanh s)^2}\big/r\right\rangle_1\left(-|0\rangle + |1\rangle\right)_2\end{array}\right)\bigg/2 =$$
$$\left(\begin{array}{l}\left(\left|0, \alpha\sqrt{1+(\tanh s)^2}\big/r\right\rangle_2 + \left|0, -\alpha\sqrt{1+(\tanh s)^2}\big/r\right\rangle_1\right)|0\rangle_2 + \\ \left(\left|0, \alpha\sqrt{1+(\tanh s)^2}\big/r\right\rangle_2 - \left|0, -\alpha\sqrt{1+(\tanh s)^2}\big/r\right\rangle_1\right)|1\rangle_2 + \end{array}\right)\bigg/2 =$$

(98)

The states are orthogonal to each other

$$\langle\Phi_-|\Phi_+\rangle = 0,$$

(99)

and they can be considered as the base elements of the output Hilbert space.

$$\left|\hat{0}_1\right\rangle = \left|\Phi_+\right\rangle_{12},$$ (100)

$$\left|\hat{1}_1\right\rangle = \left|\Phi_-\right\rangle_{12}.$$ (101)

The values of the parameter of $\delta = \pm 1$ correspond to the displacement amplitudes (60)

$$\alpha = \pm \sinh s \cosh s.$$ (102)

The same interferometer in figure 6 is used to realize operation

$$P_1(\pi) M_4^{(1)} B_{12} B_{13} B_{24} \bar{B}_{12} \left( \left| \Psi_{ab} \left( \alpha t \sqrt{1 + (\tanh s)^2} \big/ r \right) \right\rangle_1 \left| 0 \right\rangle_2 S(r) \left| 00 \right\rangle_{34} \right) \rightarrow$$
$$\left( (a+b)/\sqrt{2} \right) \Phi_+ \right\rangle_{12} + (a-b) \left| \Phi_- \right\rangle_{12} \big/ \sqrt{2}$$ (103)

provided that detector in mode 4 registered a click. The outcome can be fully recognized as outcome of Hadamard matrix

$$H = \frac{1}{\sqrt{2}} \begin{vmatrix} 1 & 1 \\ 1 & -1 \end{vmatrix}$$ (104)

acting on the state

$$H \left| \Psi_1 \right\rangle = H \left( a \left| \hat{0} \right\rangle + b \left| \hat{1} \right\rangle \right) = \left( (a+b)/\sqrt{2} \right) \left| \hat{0}_1 \right\rangle + \left( (a-b)/\sqrt{2} \right) \left| \hat{1}_1 \right\rangle,$$ (105a)

or the same on column vector of the input amplitudes

$$H \begin{vmatrix} a \\ b \end{vmatrix} = \frac{1}{\sqrt{2}} \begin{vmatrix} a+b \\ a-b \end{vmatrix},$$ (105b)

where input and output base elements are chosen from different two-dimensional Hilbert spaces.

Reverse operation can be also realized. To show it let us make use of $\alpha$ − representation of superposition of vacuum and single photon (70, 71). Consider mixing of the superposition with coherent state on UBS (85) with $\varphi = \pi$ followed by the operator $M_1^{(1)} = \left| 1 \right\rangle \left\langle 1 \right|_1$

$$M_1^{(1)} B_{12} \left( \left| 0 \right\rangle + \left| 1 \right\rangle \right)_1 \big/ \sqrt{2} \left| 0, \left( \alpha \sqrt{1 + (\tanh s)^2} \big/ r \right) \right\rangle_2 = \left( \exp \left( -\left| \alpha' \right|^2 \big/ 2 \right) \big/ \sqrt{2} \right) M_1^{(1)} B_{12}$$

$$\left( \left( 1 + \alpha' \right) \left| 0, \alpha' \right\rangle_1 + \sum_{n=1}^{\infty} \frac{(-1)^n \left( \alpha' \right)^{n-1} \left( \alpha' - \left( n - \left| \alpha' \right|^2 \right) \right)}{\sqrt{n!}} \left| n, \alpha' \right\rangle_1 \right) \left| 0, \alpha \sqrt{1 + (\tanh s)^2} \big/ r \right\rangle_2 =$$

$$M_1^{(1)} \left( \left( 1 + \alpha' \right) \left| 0 \right\rangle_1 + \sum_{n=1}^{\infty} \frac{(-1)^n \left( \alpha' \right)^{n-1} \left( \alpha' - \left( n - \left| \alpha' \right|^2 \right) \right)}{\sqrt{n!}} \left| n \right\rangle_1 \right) \left| 0, \alpha \sqrt{1 + (\tanh s)^2} \big/ r \right\rangle_2 \rightarrow$$

$$\left| 0, \alpha t \sqrt{1 + (\tanh s)^2} \big/ (Tr) \right\rangle_2$$ (106)

$$M_1^{(1)} B_{12} \left( \left( -|0\rangle + |1\rangle \right)_1 \big/ \sqrt{2} \right) \left| 0, \left( -\alpha \sqrt{1 + (\tanh s)^2} \big/ r \right) \right\rangle_2 = \left( \exp\left( -|\alpha'|^2 \big/ 2 \right) \big/ \sqrt{2} \right) M_1^{(1)} B_{12}$$

$$\left( -\left( 1 + \alpha' \right) \left| 0, -\alpha' \right\rangle_1 - \sum_{n=1}^{\infty} \frac{(\alpha')^{n-1} \left( \alpha' - \left( n - |\alpha'|^2 \right) \right)}{\sqrt{n!}} \left| n, -\alpha' \right\rangle_1 \right) \left| 0, -\alpha \sqrt{1 + (\tanh s)^2} \big/ r \right\rangle_2 =$$

$$M_1^{(1)} \left( -\left( 1 + \alpha' \right) \left| 0 \right\rangle_1 - \sum_{n=1}^{\infty} \frac{(\alpha')^{n-1} \left( \alpha' - \left( n - |\alpha'|^2 \right) \right)}{\sqrt{n!}} \left| n \right\rangle_1 \right) \left| 0, -\alpha \sqrt{1 + (\tanh s)^2} \big/ r \right\rangle_2 \rightarrow$$

$$\left| 0, -\alpha t \sqrt{1 + (\tanh s)^2} \big/ (Tr) \right\rangle$$

, (107)

где $T = t^2$, $\alpha' = \alpha \sqrt{1 + (\tanh s)^2} \big/ t$ and general sign $-$ is left out. Amplitudes of the coherent states (106. 107) coincide with input ones (87, 88) in the case of $T \rightarrow 1$. Finally, we have

$$H^{-1} \left( a \left| \hat{0}_1 \right\rangle + b \left| \hat{1}_1 \right\rangle \right) = H^{-1} \left( a |\Phi_+\rangle_{12} + b |\Phi_-\rangle_{12} \right)$$

$$\rightarrow \left( a + b \big/ \sqrt{2} \right) \left| \hat{0} \right\rangle + \left( a - b \big/ \sqrt{2} \right) \left| \hat{1} \right\rangle =$$

, (108)

$$\left( (a+b) \big/ \sqrt{2} \right) \left| 0, \alpha t \sqrt{1 + (\tanh s)^2} \big/ (Tr) \right\rangle_1 + \left( (a-b) \big/ \sqrt{2} \right) \left| 0, -\alpha t \sqrt{1 + (\tanh s)^2} \big/ (Tr) \right\rangle$$

where $H^{-1}$ is a matrix reverse to Hadamard and $H = H^{-1}$. Output state is an outcome of Hadamard matrix (104). The matrix transforms column vector of amplitudes of the hybrid qubit according to (105b).

### 3.3 Realization of Hadamard matrix between macroscopic and microscopic states

In previous subsection, realization of Hadamard gate between Hilbert spaces with base elements of different dimension is shown. Such operator is defined to be linear function between the vector spaces. Consider implementation of the Hadamard matrix between vector spaces of equal dimension, namely, between one-mode macroscopic and microscopic vector spaces. The base coherent states (87, 88) can be considered as macroscopic due to big values of their amplitudes. The basic elements of the second vector space are the microscopic states (90, 91). The same optical scheme in figure 6 is used with additional projection operator $M_1^{(n)} = |n\rangle\langle n|$ in first mode. Then, chain of transformations becomes

$$M_1^{(n)} M_4^{(1)} B_{12} B_{13} B_{24} B_{12}^{"} \left( \left| \Psi_{ab} \left( \alpha t \sqrt{1 + (\tanh s)^2} \big/ r \right) \right\rangle_1 |0\rangle_2 S(r) |00\rangle_{34} \right) \rightarrow$$

$$a \left( |0\rangle + |1\rangle \right)_1 \big/ \sqrt{2} + (-1)^n b \left( -|0\rangle + |1\rangle \right)_1 \big/ \sqrt{2} = \left( (a + (-1)^{n+1} b) \big/ \sqrt{2} \right) |0\rangle_1 + \left( (a + (-1)^n b) \big/ \sqrt{2} \right) |1\rangle_1$$

. (109)

Output state depends on number (even or odd) of the measured photons in first mode. If odd number of photons is measured, then output is exactly outcome of the Hadamard gate on column vector of input amplitudes (105b). If even number of photons is measured, then the final state is output of the matrix $U_y(Q = -\pi/2)$ that belongs to class of unitary matrices originating from Pauli matrixes when they are exponentiated [1]. So, the unitary matrix can be produced as

$$U_y(Q = -\pi/2) = iHU_x(Q = \pi),$$

(110)

where matrix corresponding to rotation around axis $y$ by angle $-\pi/2$ is composed from product of Hadamard matrix and matrix related with rotation around axis $x$ by angle $\pi$ Note only, permutation of the basic elements $\left|\hat{0}_1\right\rangle \leftrightarrow \left|\hat{1}_1\right\rangle$ in (109) enables to consider the case with even number measurement outcome as direct action of Hadamard matrix on column vector of input amplitudes.

The reverse action may be realized with help of the single-mode squeezing operator [23]

$$S_1(r) = \exp\left(r\left(a_1^{+2} - a_1^2\right)/2\right),\tag{111}$$

where $r$ is a squeezing parameter. It follows from previous works [14, 15, 20, 21], squeezed vacuum and squeezed single photon may approximate even and odd SCSs of some amplitude

$$S(r)|0\rangle \rightarrow N_+(\alpha)\big(|0,\alpha\rangle + |0,-\alpha\rangle\big),\tag{112}$$

$$S(r)|1\rangle \rightarrow N_-(\alpha)\big(|0,\alpha\rangle - |0,-\alpha\rangle\big).\tag{113}$$

Fidelity of the generated states drops with increase of amplitude $\alpha$. Increase of the SCSs size is achieved by application of photon subtraction technique to initial Gaussian states [20, 21]. We can only conjecture that squeezed superpositions of vacuum and single photon may approximate coherent states superpositions by analogy with (112, 113)

$$S(r)\big(a|0\rangle + b|1\rangle\big)/\sqrt{2} \rightarrow \big((a+b)/\sqrt{2}\big)|0,\alpha\rangle + \big((a-b)/\sqrt{2}\big)|0,-\alpha\rangle.\tag{114}$$

Increase of size of the output state may require use ща the photon subtraction procedure. Transformation (114) may be considered as reverse action of Hadamard matrix. Analysis of the conversions (114) requires separate investigation.

## 4. Realization of the elementary gates in realistic scenario

We have considered implementation of elementary gates in conjecture of ideal projective measurement. As APD cannot distinguish number of incoming photons it induces additional contribution of higher-order terms of the state (61). It leads to decrease of unity fidelity of the generated states. Let us discuss influence of real APD on the fidelity of the generated states. Single photon must be registered by APD in mode 4 for implementation of the gates. Therefore, small values of the squeezing parameter $s \ll 1$ must be taken to ensure dominance of the states $|0\rangle$ and $|1\rangle$ over other states in distribution of the TMSV. Contribution of the higher order number states can be negligible in the case of $s \ll 1$. When APD registers click in mode 4, it implies with almost unity probability that single photon was measured in the case of $s \ll 1$. For example, figure 7 shows plots of relations of single photon probability to probabilities of higher order terms $P_1(\delta=1,s)/P_k(\delta=1,s)$ with $k$ from 2 up to 7 in dependency on squeezing parameter $s$ for the case of $\delta=1$. The case corresponds to construction of Hadamard gate with hybrid (103) and microscopic (109) states. The same reasoning is applicable to consideration of control-sign gate. Thus, smallness of the squeezing parameter $s$ guaranties almost ideal fidelity of generated states.

Consider reverse implementation of the Hadamard gate (106-108) with APD. Projection operator on the displaced single photon is substituted by APD followed by the UBS. To decrease influence of the terms with single photon more $n > 1$ in (106, 107) on fidelity of the output state, we can choose some value of the displacement amplitude $\alpha'$. The choice of the value can be done with help of plot in figure 4(b) with $\alpha=1$.

Consider realization of the projective measurement in mode 1 with APD. Coherent states with amplitudes $\pm \alpha\sqrt{1+\tanh^2 s}/r$ go towards APD in mode 1. Amplitudes of the coherent states are large modulo. Therefore, we need to pass the states through absorbing medium to decrease their amplitudes. Suppose the absorbing medium transforms the coherent states as $|0, \pm\alpha\rangle \rightarrow |0, \pm\alpha/T\rangle$, where T is an absorbing coefficient. The parameter can be chosen in such a way the absolute amplitude of output states to be close to zero. The coherent states may be approximated by superposition $|0, \pm\alpha\rangle \approx |0\rangle \pm \alpha|1\rangle$. Then, if APD registers something, then it means registration of single photon with large probability. It may assure accomplishment of projection measurement onto single photon

## 5. Conclusion

We introduced transformation matrix from one set of the displaced number states to other with different displacement amplitude. Analytical expressions of the matrix elements are derived. Matrix elements connecting number states with displaced number ones are analyzed in details. The transformation matrix is a key moment for introduction of $\alpha$-representation of the state. The pure state in its $\alpha$-representation is obtained by multiplication of the transposed transformation matrix on column vector of state amplitudes and presents decomposition of the state in terms of the displaced number states with displacement amplitude $\alpha$. Analytical expressions of the $\alpha$-representation of TMSW and superposition of vacuum and single photon are derived. Derivation of $\alpha$-representation of superposition of vacuum and single photon stems from main rule. Conclusion of the $\alpha$-representation of TMSW is not trivial and requires additional mathematical transformations with matrix elements.

The treatment with $\alpha$-representation of the states enables to introduce method of extraction of displaced number states from part of the correlated system to generate the target state. Such approach sounds attractive as it opens new possibilities to manipulate the characteristics of the quantum states. Reduction postulate [1] tells that if a measurement is performed on a portion of a composite system, the output state of unmeasured part of the correlated system strongly depends on the result of the measurement. It provides an alternative mechanism to effectively achieve nonlinear effect comparable with that of nonlinear media. Moreover, as photon subtraction is achieved with zeroth amplitude of the displacement, it can be considered as partial case of extraction of displaced photon. Nonzero amplitude of the extracted displaced photon enables to get information in different coding. In particular, extraction of the displaced $n$ number state in one of two modes of the TMSV enables to realize creation operator with displacement amplitude (60) in power $n$ acting on vacuum state. The projective operator on displaced number state in quantum optics is realized with help of highly transmissive beam splitter followed by the projection operator on the number state.

Usefulness of the $\alpha$-representation of the states manifests in developing new mechanism of manipulation by quantum information. The mechanism is based on symmetry properties of the matrix elements, in particular TMSV, in regard to sign change of the displacement amplitude. Mach-Zehnder interferometer with additional inserted unbalanced beam splitters is considered for building of elementary one-qubit and two-qubit operations with help of the $\alpha$-representation of the TMSV. Control qubit is a superposition of coherent states of large amplitude able to implement displacement operation on the target qubit. Coherent qubit simultaneously shifts the target qubit on negative and positive values on phase plane. Shift on the phase plane is turned out to be required negligible. Successful registration of the single photon leads to transformation of quantum information corresponding to elementary gates. The gates are probabilistic and conditioned by successful registration of one click in auxiliary mode. It requires minimal number of resources and substantially increases success probability of the operation. We have shown one APD can be used in the scheme. The treatment is exact and no any approximations are used that increases its importance. On the other hand, it requires separate investigation to adjust the gates with different base elements to quantum information protocols. It may carry as benefits and drawbacks. Problem of applicability of such transformations to quantum computing is open.

**Acknowledgement**

The work was supported by program of development of personnel of Southern Ural State University.

## LIST OF FIGURES

**Figure 1(a-f)**
Probability distribution of the single photon state over displaced number states with different values of the displacement amplitude.

**Figure 2(a-f)**
Probability distribution of the three-photon state over displaced number states with different values of the displacement amplitude.

**Figure 3(a-f)**
Probability distribution $P_n(\delta, r)$ of the two-mode squeezed vacuum over correlated displaced number states in neighboring modes for different values of the squeezing parameter $r$ and parameter $\delta$ proportional to the displacement amplitude.

**Figure 4(a-f)**
Probability distribution $P_{n+}(\alpha)$ of superposition of vacuum and single photon over displaced number states for different values of the displacement amplitude $\alpha$.

**Figure 5(a-f)**
Probability distribution $P_{n-}(\alpha)$ of superposition of vacuum and single photon over displaced number states for different values of the displacement amplitude $\alpha$.

**Figure 6**
The optical scheme is used for realization of one- and two-qubit operations. Control coherent qubit starts in mode 1, splits onto two modes in UBS $B_{12}^{"}$, again combines on UBS $B_{12}$ and interacts with TMSV in the interim between the UBSs. TMSV occupies modes 3 and 4. Target qubit is initially contained in the TMSV. Interaction of the control qubit and TMSV occurs simultaneously on two UBSs $B_{13}$ and $B_{24}$. Registration of the single photon in mode 4 guaranties generation of output state of the gate.

**Figure 7**
Dependencies of relations of probability of displaced single photon to the probabilities of higher-order terms of TMSV $P_1(\delta = 1, s) / P_{k+1}(\delta = 1, s)$ (curve $k$), where $k$ is varied from 1 up to 6, on $s$.

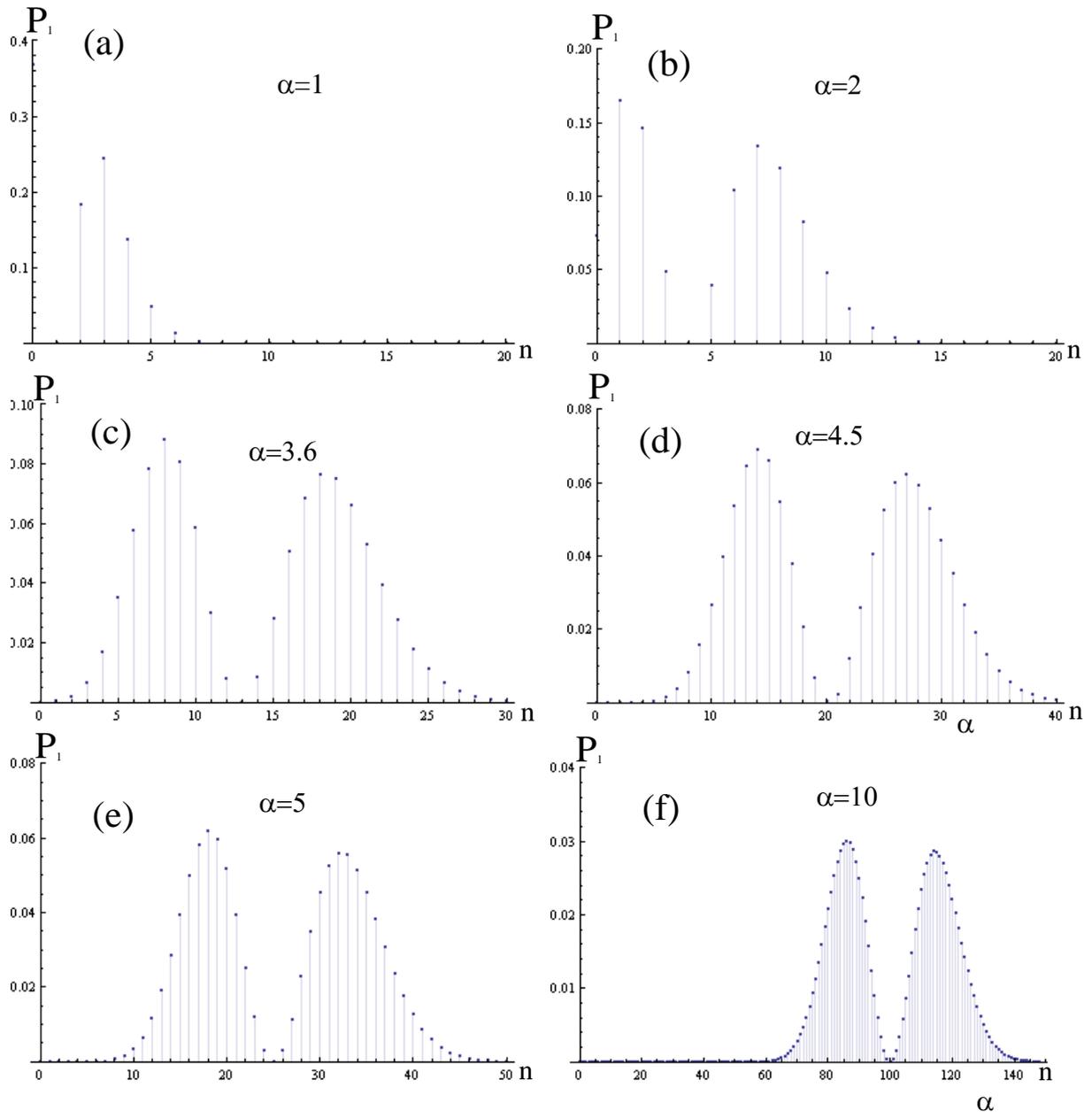

**Figure 1(a-f)**

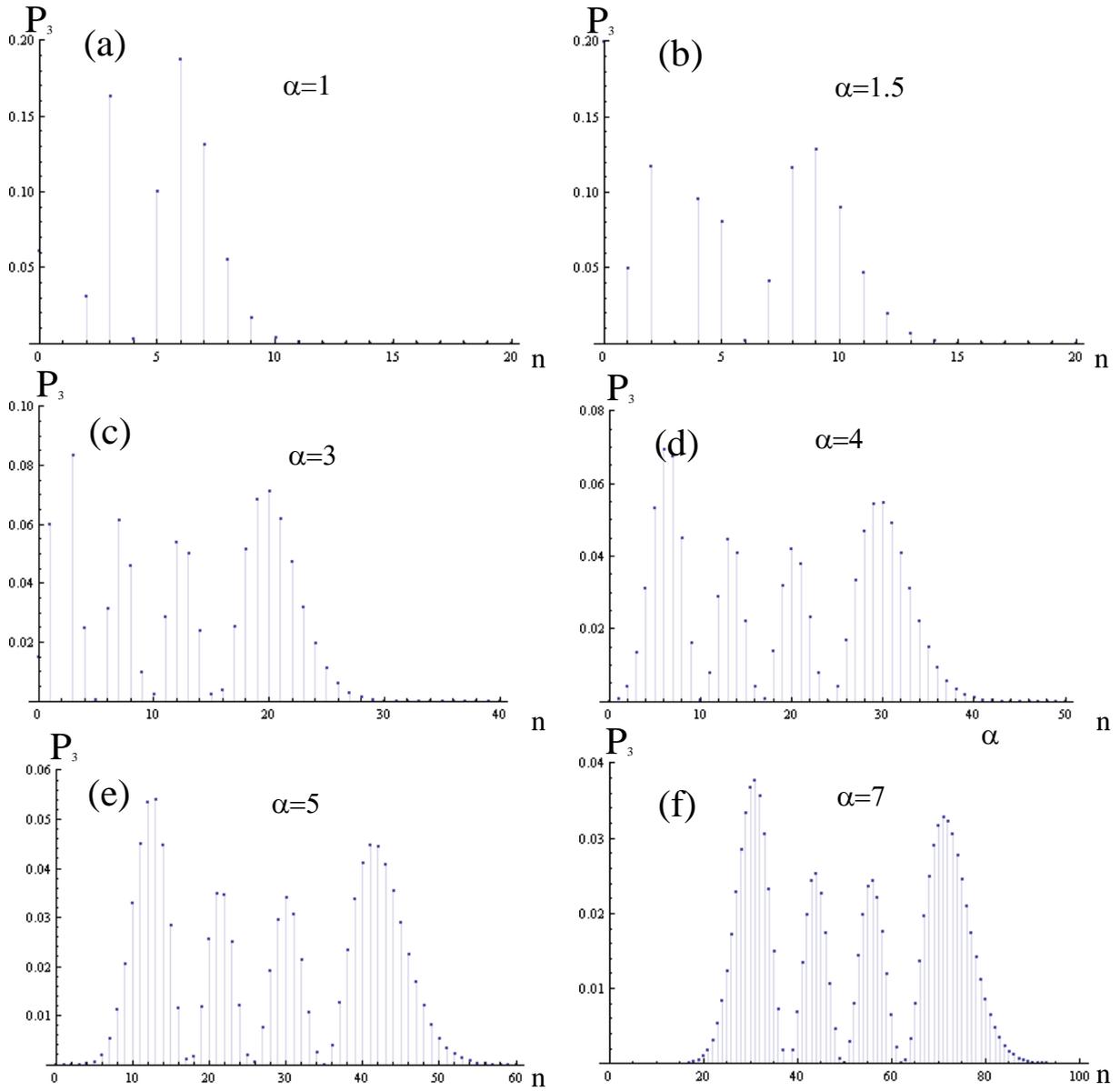

**Рисунок 2(a-f)**

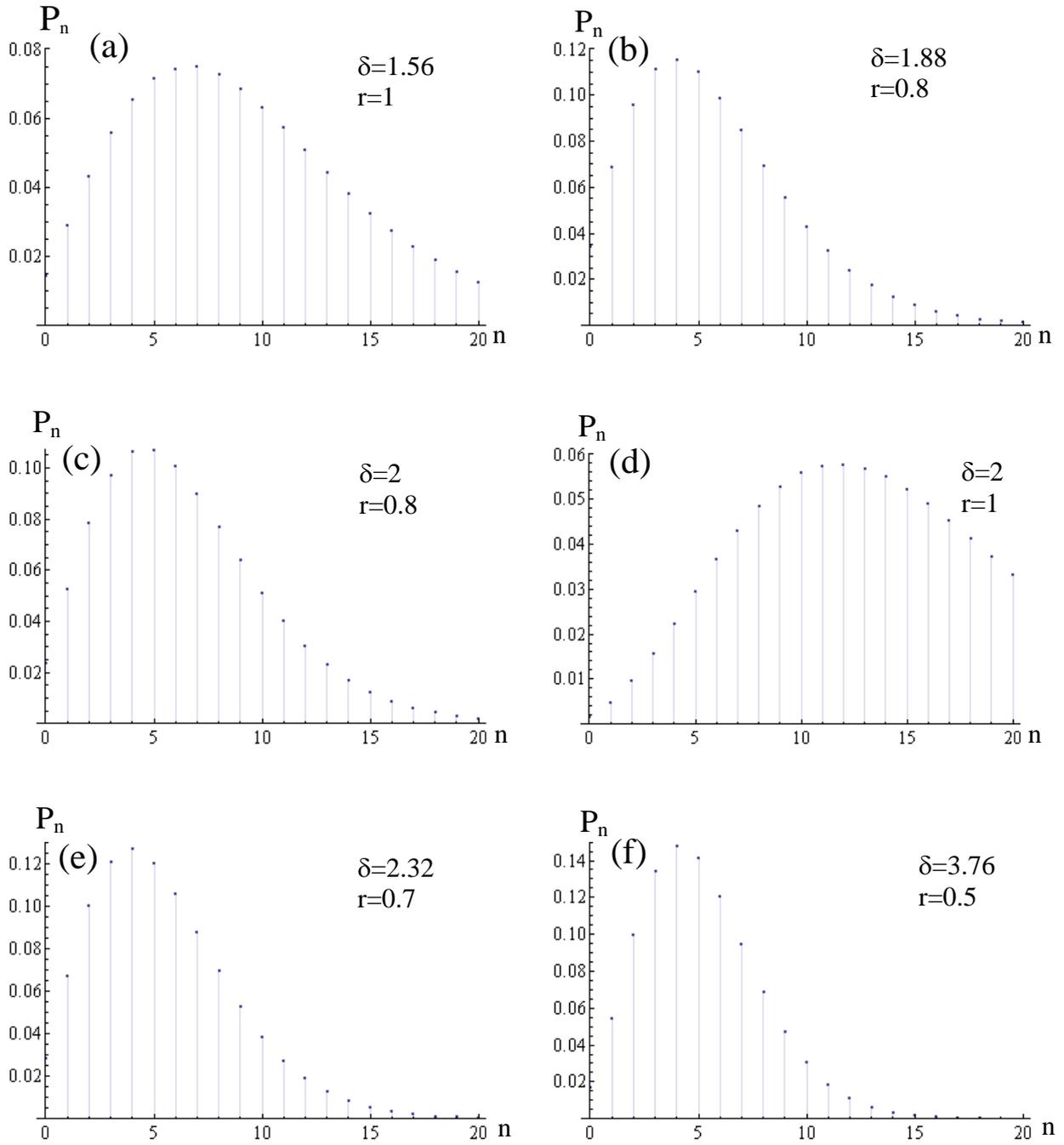

**Figure 3(a-f)**

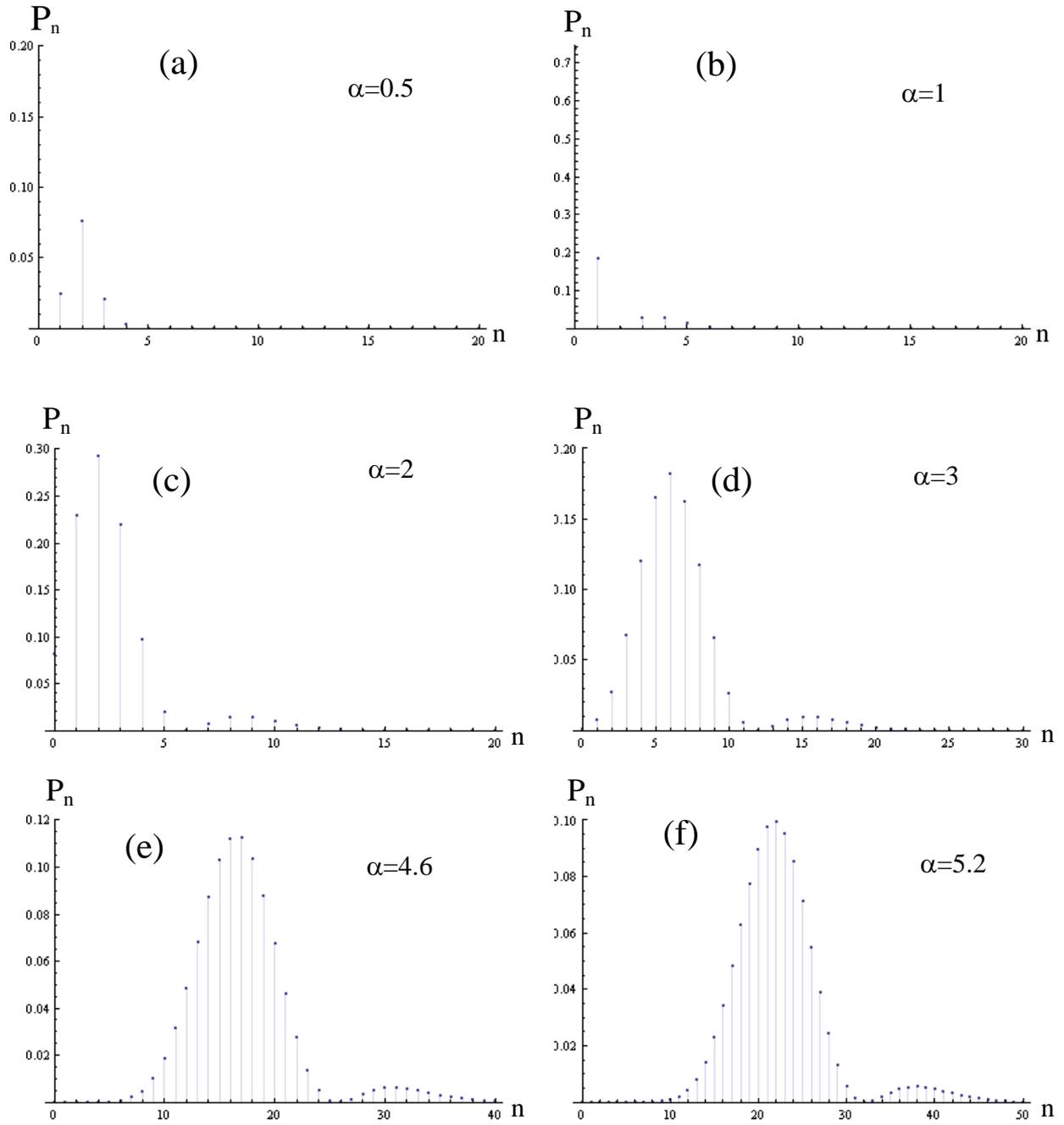

**Figure 4(a-f)**

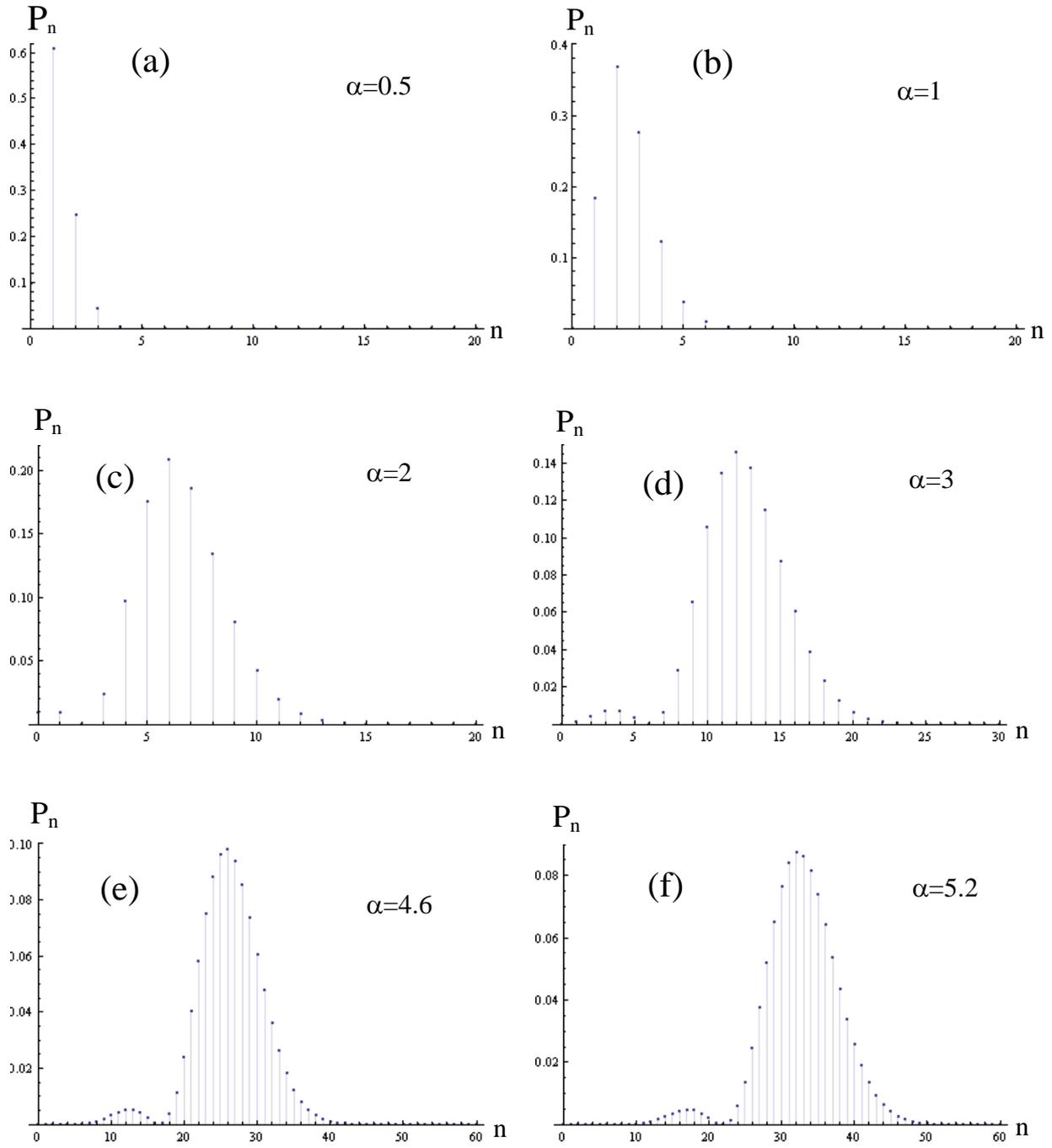

**Figure 5(a-f)**

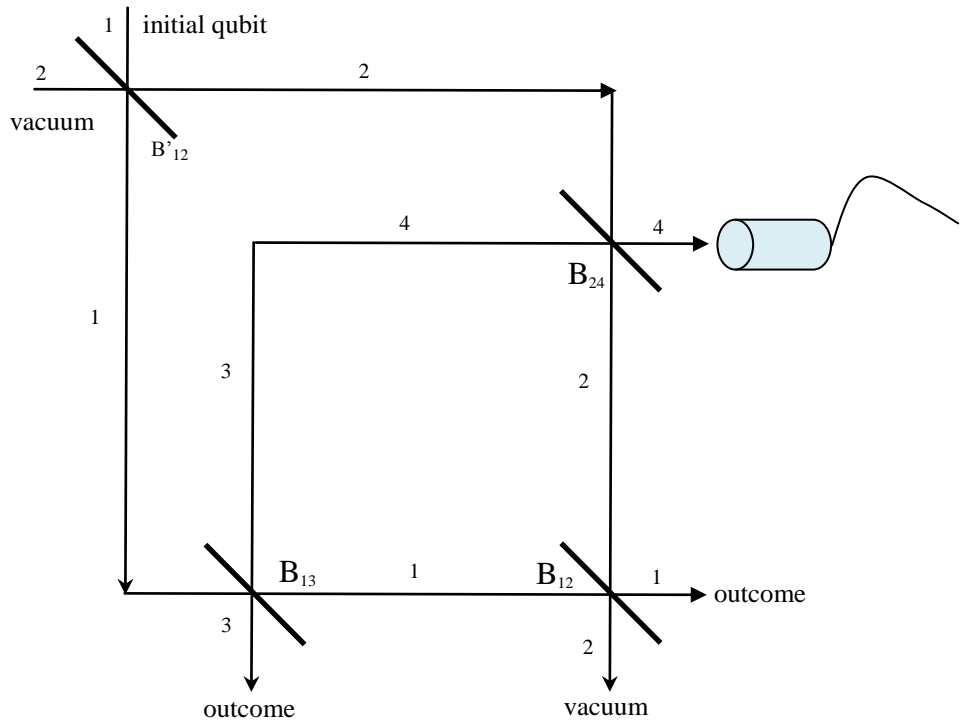

**Figure 6**

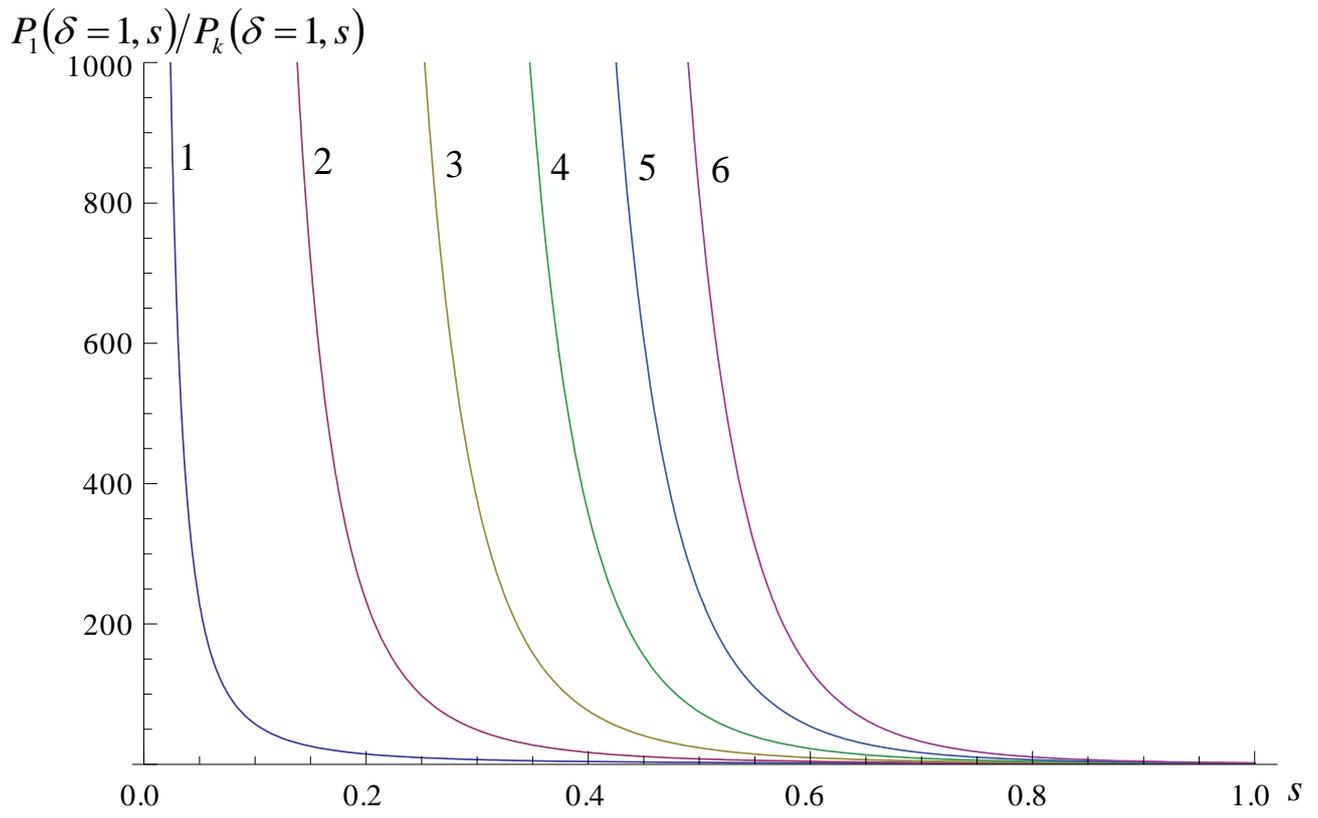

**Figure 7**